\documentclass{optica-article}
\journal{opticajournal} % for journals or Optica Open
\usepackage{xspace}
\usepackage{gensymb}
\usepackage{graphicx}
\usepackage{caption}
\usepackage{subcaption}
\usepackage{url}
\usepackage{hyperref}
\usepackage{tabularx}
\usepackage{lineno}
\usepackage{float}

\newcommand{\etal}{\textit{et al.\@}\xspace}

\newcommand{\invivo}{\textit{in vivo}\xspace}

\newcommand{\enface}{\textit{en face}\xspace}

\newcommand{\um}{{$\mu$m}\space}

\newcommand{\bmat}[1]{\mathbf{#1}\xspace}
\newcommand{\bJ}{\bmat{J}\xspace}
\newcommand{\Jin}{{\bJ_{\mathrm{in}}\xspace}}
\newcommand{\Jout}{{\bJ_{\mathrm{out}}\xspace}}

\begin{document}
\title{Polarization-artifact reduction and accuracy improvement of Jones-matrix polarization-sensitive optical coherence tomography by multi-focus-averaging based multiple scattering reduction}

\author{
Lida Zhu,\authormark{1} 
Shuichi Makita,\authormark{1} 
Junya Tamaoki,\authormark{2} 
Yiqiang Zhu, \authormark{1}
Pradipta Mukherjee, \authormark{1}
Yiheng Lim,\authormark{1} 
Makoto Kobayashi,\authormark{2} 
and Yoshiaki Yasuno\authormark{1,*}}

\address{\authormark{1}Computational Optics Group, University of Tsukuba, Tsukuba, Ibaraki, Japan\\}
\address{\authormark{2}Department of Molecular and Developmental Biology, Institute of Medicine, University of Tsukuba, Tsukuba, Ibaraki, Japan\\}
\email{\authormark{*}yoshiaki.yasuno@cog-labs.org} %% email address is required
\homepage{https://optics.bk.tsukuba.ac.jp/COG/} %% author's URL, if desired

\begin{abstract}
Polarization-sensitive optical coherence tomography (PS-OCT) is a promising biomedical imaging tool for differentiation of various tissue properties. 
However, the presence of multiple-scattering (MS) signals can degrade the quantitative polarization measurement accuracy. 
We demonstrate a method to reduce MS signals and increase the measurement accuracy of Jones matrix PS-OCT.
This method suppresses MS signals by averaging of multiple Jones matrix volumes measured using different focal positions. 
The MS signals are decorrelated among the volumes by focus position modulation and are thus reduced by averaging.
However, the single scattering signals are kept consistent among the focus-modulated volumes by computational refocusing.
We validated the proposed method using a scattering phantom and a postmortem medaka fish.
The results showed reduced artifacts in birefringence and degree-of-polarization uniformity measurements, particularly in deeper regions in the samples.
This method offers a practical solution to mitigate MS-induced artifacts in PS-OCT imaging and improves quantitative polarization measurement accuracy.
\end{abstract}

%\keywords{Polarization sensitive, optical coherence tomography, multiple scattering, tunable lens, computational refocusing}

\section{Introduction}
\label{sec:intro} 
Optical coherence tomography (OCT) is a non-invasive and label-free imaging modality that visualizes the three-dimensional (3D) morphology of a sample using the backscattering intensity as a contrast source \cite{Huang1991Science}.
Polarization-sensitive OCT (PS-OCT), a functional extension of OCT, provides additional contrasts that are sensitive to a diverse range of tissue properties\cite{deBoer_BOE2017, Baumann_2017}. 
One example of these contrasts is the birefringence, which is effective for identification of fibrous tissues such as muscle and collagen\cite{deBoer_OL1997, Nadkarni_2007, Miyazawa2009OpEx, Lim2011BOE}. 
Another example is the degree-of-polarization uniformity (DOPU), which measures the polarization randomness. 
The DOPU is sensitive to the polarization scrambling property of the sample and has been used in detection and diagnosis of the retinal pigment epithelium layer \cite{G_tzinger_2008} and the choroid \cite{Miura2022SciRep_VKH, Miura2022SciRep_Normal} in the human retina. 
These contrasts in PS-OCT enable comprehensive investigations of tissue properties in biological studies\cite{Yang_JBio2022, Lichtenegger_Bioeng2022} 

Typically, both OCT and PS-OCT rely on use of single-back-scattering (SS) signals to reconstruct the images of a sample\cite{Fercher_2003}. 
However, in practice, OCT and PS-OCT images both deteriorate by the multiple-scattering (MS) signals, especially when the sample exhibit strong scattering.
In such cases, the SS signal at a specific depth is overlaid by MS signals originating from different depths, and this effect causes contrast degradation of the image. 
Additionally, these MS signals can interfere with polarization measurements. 
For example, Lichtenegger \etal measured zebrafish using a Jones-matrix PS-OCT (JM-OCT) and noticed erroneously increased birefringence and reduced DOPU values\cite{Lichtenegger_BOE2022}.

To mitigate the image degradation caused by the MS signals, several wavefront shaping-based methods have been demonstrated. 
In these methods, the probe beam’s wavefront is modulated using a control device such as a spatial light modulator (SLM) \cite{badon_SA_2016, Wojtkowski_stoc2019}, a deformable mirror (DM) \cite{liu_BOEB_2018}, or an electrical tunable lens (ETL) \cite{LZhu2023BOE}.
Because the modulation process decorrelates the MS and SS signals, only the MS signals can be eliminated by the post-acquisition signal processing methods.
Although the methods noted above improved the contrast of OCT intensity images, their effects on PS-OCT imaging have not been demonstrated to date. 

In this study, we modify and adapt our previously demonstrated wavefront shaping-based MS signal reduction method, the multi-focus averaging (MFA) method \cite{LZhu2023BOE}, for application to polarization imaging in JM-OCT.
In this method, we simultaneously modulate four OCT images that form a measured Jones matrix using an ETL.
Specifically, multiple JM-OCT volumes are acquired sequentially with different defocus configurations. 
The defocus is corrected by a computational refocusing process, and this process decorrelates the SS and MS signals.
Subsequent complex averaging of the volumes reduces the MS signals while retaining the SS signals.
The proposed method was validated by measurements of a phantom and a postmortem medaka fish.
We found that the MS signals caused errors in the DOPU and birefringence measurements, particularly in the deeper regions in which the MS signals are prominent, and also found that these errors can be mitigated using our MFA method. 

\section{Principle and implementation}
\subsection{Principle}
\label{sec:principle:theory}
The JM-OCT measures the cumulative Jones matrix of the light that is backscattered from a sample and multiple contrasts can then be computed from the measured Jones matrix.  
The measured Jones matrix $\bJ_m$ at a depth $z$ can be represented by
\begin{equation}
	\label{eq:jm}
	\bJ_m (x,y;z) = \Jout \bJ_s(x,y;z) \Jin,
\end{equation}
where $x$ and $y$ are the lateral positions, and $\Jin$ and $\Jout$ represent the Jones matrices of the illumination and collection paths of the system, respectively.
$\bJ_s$ is the round-trip cumulative Jones matrix of the sample within a small depth region around a depth of $z$. 
Notably, $\bJ_m$ has four entries that correspond to four complex OCT images (or volumes) acquired using the four polarization channels of the JM-OCT system.

We then apply our previously demonstrated MFA method to each of the entries (i.e., to each of the complex OCT volumes) in the measured Jones matrix $\bJ_m$.
Because the principle of MFA can be found in detail elsewhere \cite{LZhu2023BOE}, we present only a brief summary of the method here.
Assuming that the depth $z$ in the OCT image experiences a defocus amount of $z_d$, the complex OCT signal of the $j$-th entry of $\bJ_m$ can then be written as
\begin{equation}
	\label{eq:defocusedSignal}
	S^j(x,y;z, z_d) = S^j_{SS}(x,y;z, 0) * \exp \left[i \phi(x,y;z, z_d) \right]+ S^j_{MS}(x,y;z, z_d),
\end{equation}
where $*$ denotes the convolution operation, and $S^j_{SS}$ and $S^j_{MS}$ denote the SS and MS signal components, respectively.
$\phi$ is a defocus-dependent quadratic phase function and the phase term convolved with the SS component term represents the defocus.

By applying a computational refocusing process to the complex OCT signal of the $j-$th entry, the signal then becomes
\begin{equation}
	\label{eq:refocusedSignal}
	{S^j}'(x,y;z, z_d) = S^j_{SS}(x,y;z, 0) + S^j_{MS}(x,y;z, z_d) * \exp \left[-i \phi(x,y;z, z_d) \right]. 
\end{equation}
As shown in Eq.\@ (\ref{eq:refocusedSignal}), the refocused SS signals $S^j_{SS}$ remain consistent, regardless of the different defocus amounts.
In contrast, the MS lights follow practically random paths and are thus altered by the defocus.
Therefore, the MS signals cannot remain consistent with variations in the defocus amount, even after computational refocusing.
Therefore, averaging of the multiple focus-modulated OCT signals after application of the refocusing process can reduce the MS signals while preserving the SS signals as follows: 
\begin{equation}
	\label{eq:mfaSignal}
	\overline{{S^j}'}(x,y;z) = S^j_{SS}(x,y;z, 0) + \frac{1}{N} \sum_{i=0}^{N-1} S^j_{MS}(x,y;z, z_{d, i}) * \exp \left[-i \phi(x,y;z, z_{d,i}) \right],
\end{equation}
where $\overline{{S^j}'}$ is the averaged OCT signal, $N$ is the number of defocus-modulated OCT volumes for averaging, and $z_{d, i}$ denotes the defocus amount of the $i-$th acquisition. 

Because all four entries, i.e., the four complex OCT volumes contained in a volumetric measured Jones matrix, are assumed to experience equal defocus, the same computational refocusing process can be applied to each of the four entries. 
The OCT intensity, the birefringence, and the DOPU are then computed from a new Jones matrix constructed using the averaged OCT signals.
The implementation of this method will be described in detail in Section \ref{sec:implement:jmmfa}

\subsection{Implementation}
\subsubsection{JM-OCT setup}
\label{sec:implement:system}
A custom-built JM-OCT system was used in this study.
%\yascom{Fill the scanning band width. It is not necessarily be the -10 dB width, but don't forget to correct the description of the width-definition.}
The system uses a wavelength-swept light source (AXP50124-8, Axsun Technologies, MA) with a wavelength centered at 1,310 nm and a scan width of approximately 106 nm when measured at -10 dB width.
The axial resolution was measured to be 14 \um in tissue.
The A-line rate of this system is 50 kHz. 
Use of an objective with an effective focal length of 36 mm (LSM03, Thorlabs, NJ) and an incident beam diameter of 3.49 mm on the objective provides a lateral resolution of 18 \um and a depth-of-focus (DOF) of 0.36 mm in air.
An ETL (EL-10-30-CI-NIR-LD-MV, Optotune, Switzerland) was integrated into the sample arm to modulate the defocus for the MFA method.
Further details about the system with the exception of the ETL part can be found in Refs.\@ \cite{li_BOEB_2017, miyazawa_BOEB_2019}, and details of the ETL implementation can be found in Section 4.4 of Ref.\@ \cite{LZhu2023BOE}. 

\subsubsection{MFA signal processing for JM-OCT	}
\label{sec:implement:jmmfa}
The JM-OCT-based MFA measures multiple volumetric Jones matrices with the focus at different depth positions, i.e., with different defocus amounts.
These volumetric Jones matrices consist of four complex OCT volumes that correspond to the entries in the Jones matrix.
The volumetric Jones matrices are processed using a signal processing flow that consists of bulk-phase error correction, computational refocusing, inter-volume axial shift correction, and inter-volume phase offset correction, and the resulting matrices are then averaged to reduce the MS signals.
Multi-contrast OCT images, including the OCT intensity, DOPU, and birefringence are computed from the averaged Jones matrix volume.
This process is similar to our previous MFA method but has been modified to deal with the Jones matrix signals.
The processing details are described in the following.

First, for each acquired volumetric Jones matrix, we estimate and correct the bulk phase errors using a smart-integration-path method \cite{oikawa_BOEB_2020}. 
The bulk phase error is estimated from a single entry in the Jones matrix and the errors in all entries are then corrected using this estimate.

Second, a computational refocusing process was applied to correct the defocus. 
We estimated the defocus amount at each depth from a single entry in the volumetric Jones matrix and then corrected the defocus for all entries using this defocus amount.
Details of the refocusing process have been presented in Section 2.2 of Ref.\@ \cite{zhu_BOE_2022}. 

The deformation of the liquid lens in the ETL causes an axial shift in the image.
In the third step, this shift is estimated and corrected using a sub-pixel linear intensity cross-correlation method \cite{hillmann_OE_2012}. 
The axial shift is estimated from a single entry in the volumetric Jones matrix and this shift is used to correct all four entries. 
Details of the process for estimation for a single entry are given in Section 2.2.2 in Ref.\@ \cite{LZhu2023BOE}. 

In the fourth step, the inter-volume phase offsets are estimated using all four entries (i.e., the volumes), while they were estimated from a single entry in our previous MFA method.
In this estimation process, one volumetric Jones matrix among the Jones matrices that were obtained with the different defocus amounts is designated as a reference.
The product of the reference volume and the complex conjugate of another volume, which is regarded as the target volume, is computed for each entry.
These products are averaged in complex form along the depth direction and are then averaged over the four entries, thus allowing a complex \enface map to be obtained. 
The phase of the \enface map represents the relative phase offset of the target volume with respect to the reference volume. 
The estimated phase offset is used to correct the phase offsets of all four entries in the target volume. 

After all the aforementioned corrections have been performed, the corrected Jones matrix volumes are averaged in complex form, and multi-contrast OCT images, including the OCT intensity, the DOPU, and the birefringence, are computed from the averaged Jones matrix.
The intensity image is the intensity average of all four entries.
In our implementation, the DOPU is computed by applying Makita's noise correction \cite{Makita_OL2014} using a spatial kernel of 3 $\times$ 3 pixels (17.5 \um in the lateral direction $\times$ 21.7 \um in the axial direction). 
%%%
To better visualize the scattering tissue of particular interest, in our image formation, the DOPU value was mapped to a rainbow color map and converted to hue-saturation-lightness color space. 
Then, the saturation value is replaced with the log-scale OCT signal intensity. 
As a result, the low intensity background is shown as gray. 
%%%
For the birefringence computation (i.e., the local phase retardation computation), we first compute the local Jones matrix with an axial separation of 8 pixels (57.9 \um in tissue).
The birefringence is then computed using the maximum \textit{a posteriori} (MAP) birefringence estimator \cite{Kasaragod_OE2014, Kasaragod_BOE2017}.
The estimation kernel used here is a single pixel, i.e., the estimator morphed the single value of the measured birefringence into its maximum likelihood estimation using a numerically obtained likelihood function and the measured effective signal-to-noise ratio (SNR).
Details of the image formation processes for these contrasts can be found in Section 2.3 of Ref.\@ \cite{zhu_BOE_2022}. 

\subsection{Validation study design}
\label{sec:validation}
\subsubsection{Samples}
\label{sec:validation:sample}
\begin{figure}
	\centering\includegraphics[width=11cm]{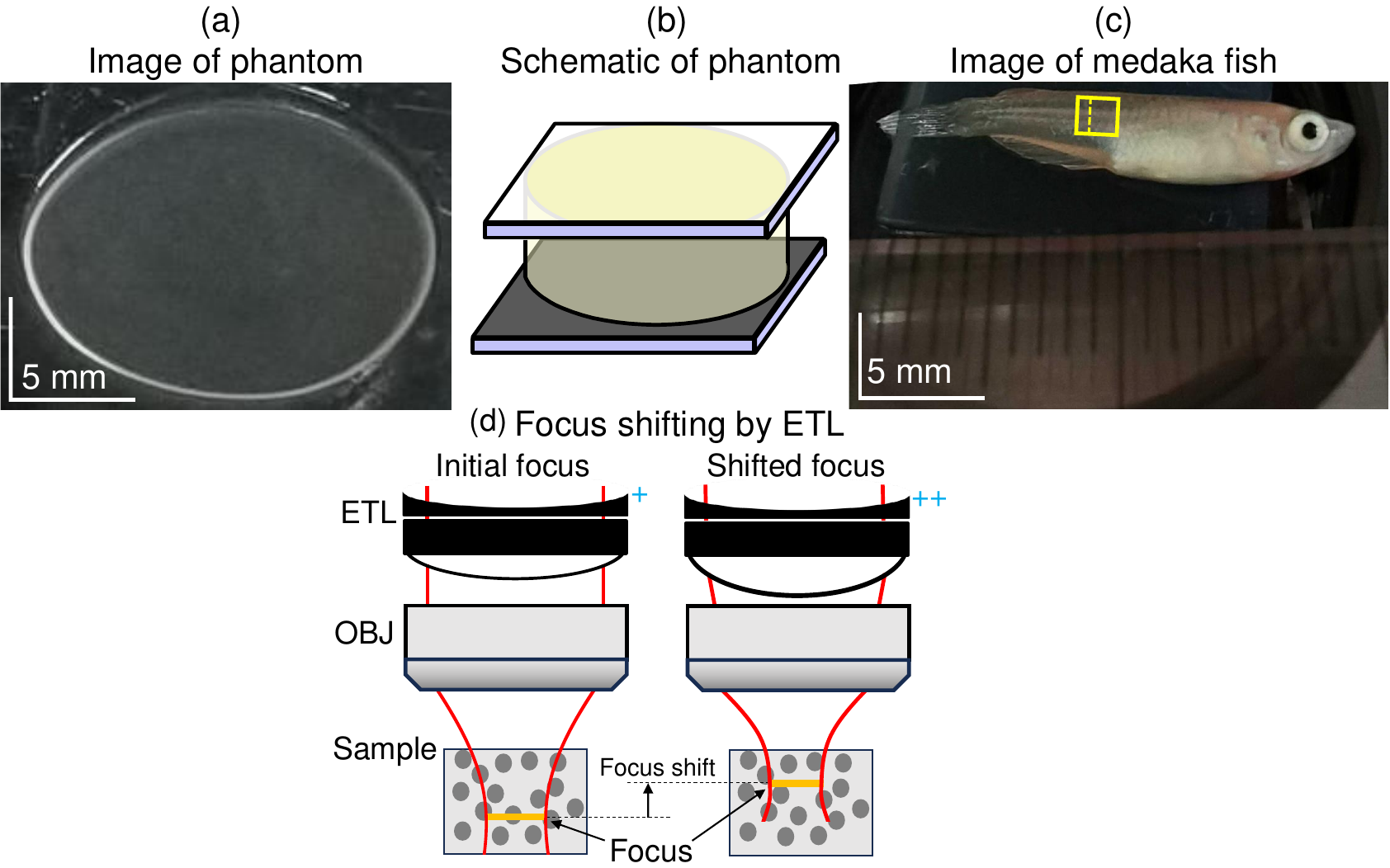}
	\caption{
		A scattering phantom sample and a postmortem medaka fish were used to validate the MS-signal rejection method.
		(a) and (b) show a photograph and a schematic of the scattering phantom, respectively. 
		This phantom is composed of a scattering layer sandwiched between two glass plates, where the scattering layer is a mixture of polystyrene microparticles with a diameter of 10 \um and ultrasound gel.
		The particle density is 0.2 g/mL.
		(c) shows a photograph of the postmortem medaka fish, where the yellow box and the dashed line indicate the measurement area and the location of the cross-sectional images shown in Figs.\@ \ref{fig3:fish_int}--\ref{fig5:fish_biref}, respectively. 
		(d) schematically illustrates the modulation of the focal position by an ETL.
		ETL, an electrical tunable lens, and OBJ, an objective.
	}
	\label{fig:samples}
\end{figure}
To validate the proposed method, we measured two sample types.
The first was a non-birefringent scattering phantom that was similar to the phantom used in our previous MFA study \cite{LZhu2023BOE}.
A photograph of this phantom is shown in Fig.\@ \ref{fig:samples}(a).
As the phantom schematic shows [Fig.\@ \ref{fig:samples}(b)], the phantom consists of a glass slip at the top, a scattering layer in the center, and another glass slip with black tape at the bottom.
The scattering medium of this phantom was composed of a mixture of 10\%-concentration polystyrene microparticles (diameter of 10 \um, 72986-10ML-F, Sigma-Aldrich, St Louis, MO) and transparent ultrasound gel (Pro Jelly, Jex, Japan). 
The particle density of the resulting phantom is 0.2 g/mL. 

The second sample is an adult postmortem medaka fish, which contains birefringent tissues such as muscle. 
A photograph of the sample fish is shown in Fig.\@ \ref{fig:samples}(c).
Before conducting the measurements, we anesthetized the fish using tricaine and then sacrificed it by placing it on ice for 2 min. 
We placed the fish in a petri dish and immersed it in a saline solution to perform the measurements.
This protocol follows the animal experiment guidelines of the University of Tsukuba and is approved by the Institutional Animal Care and Use Committee of the University of Tsukuba. 

\subsubsection{Measurement protocols}
\label{sec:validation:protocol}
During the measurements of both the phantom and the medaka fish, we acquired seven volumes with different focal positions for the averaging process. 
The focal position of the probe beam in the sample was modulated by the ETL, as schematically illustrated in Fig.\@ \ref{fig:samples}(d). 
We set the interval between the focal positions at 0.12 mm, which represents approximately one-third of the DOF. 
It results in an overall focus shifting distance of 0.72 mm for the set of seven acquisitions.
We determined these parameters based on an experimental optimization process that is described in Section 4.2 of Ref.\@ \cite{LZhu2023BOE}. 
The images obtained by this method are referred to as the MFA images in the later sections.

We also acquired another set of volumes without focus shifting for comparison purposes.
We processed these non-focus-shifting volumes using the same processing flow that was described in Section \ref{sec:implement:jmmfa}, and the resulting images are referred to here as the single focus averaging (SFA) images.

In addition, images that were obtained from a non-averaged but defocus-corrected single Jones matrix were computed and are referred to here as single-acquisition images.
The single-acquisition, SFA, and MFA methods are compared qualitatively and quantitatively. 

During the measurements, the focal positions were set at depths of more than 1 mm below the sample surface. 
The lateral imaging area was 1.5 mm $\times$ 1.5 mm, and this area was sampled using 256 $\times$ 256 A-lines.
This scan protocol results in an isotropic lateral pixel separation of 5.86 \um $\times$ 5.86 \um, which is approximately one-third of the lateral optical resolution. 
The total acquisition time for the seven volumes was approximately 20 s. 

\subsubsection{Signal evaluation metrics}
\label{sec:validation:metric}
In this study, we used the following metrics to compare among the images obtained by single-acquisition, SFA, and MFA methods. 

In the phantom measurement results, we computed the following metrics. 
For depth-dependent analysis, the means of intensity, DOPU, and birefringence along the depth were computed and plotted as the depth profiles. Higher mean DOPU values that are closer to 1 and lower mean intensity may indicate better noise reduction, where the noise includes MS signals and standard noises. 
For \enface images comparison, the modes of histograms were computed from the \enface images at two different depths.
Higher mode of DOPU may indicate a better noise reduction 

In the medaka fish measurement results, we computed the following metrics. 
For tissue-depth-dependent analysis, four region-of-interests (ROIs) at different depths of the skin and muscle regions were selected from a cross-sectional image. 
For each ROI, the means, medians, upper and lower quartiles of the DOPU and birefringence were computed and plotted in box plots. 
Higher mean and median values of DOPU may indicate a better noise reduction. 
For both skin and muscle ROIs, the differences of the mean DOPU and mean birefringence values between the ROIs in the upper and lower half of the fish were computed and denoted as ``upper-lower difference.'' 
Here the ``upper'' and ``lower'' mean the upper and lower part of the image, which corresponds to the right and left sides of the fish. 
Because of the anatomical symmetricity of the fish, if ideal signals were obtained, the mean values should be the same for either the skin and muscle ROIs. 
Namely, smaller difference indicates a better reduction of the MS-induced artifacts. For the tissue-dependent analysis at the same depth, the birefringence contrast was computed from two manually selected ROIs in the \enface birefringence images at a deep depth. 
Higher contrast may indicate a better artifact reduction.

One of the major disturbances of quantitative analysis is measurement noise, and it also affects the DOPU and birefringence measurements \cite{Makita_OL2014, Kasaragod_BOE2017}. 
Hence, we applied an intensity threshold of +5 dB to exclude the low SNR region from the quantitative analysis. 
Here the 0 dB was defined as the mean intensity at the air of the single-acquisition images. 
This thresholding improves the accuracy of quantification by excluding the low intensity pixels.
However, it should be noted that this exclusion may occasionally select highly scattering tissues and remove low scattering tissues. 
This potentially biased selection of tissue types may affect the performance of the evaluation. 
So, we should be careful to avoid the overgeneralization of the results and conclusions.

\section{Results}
\label{sec:result}
Note that all OCT intensity images presented in this paper are displayed in dB scale, where 0 dB refers to the noise floor in the air region in the single-acquisition images.

\subsection{Scattering phantom}   
\label{sec:result:phantom}
\subsubsection{OCT intensity and DOPU of the phantom}   
\label{sec:result:phantom_dopu}
\begin{figure}
	\centering\includegraphics[width=13cm]{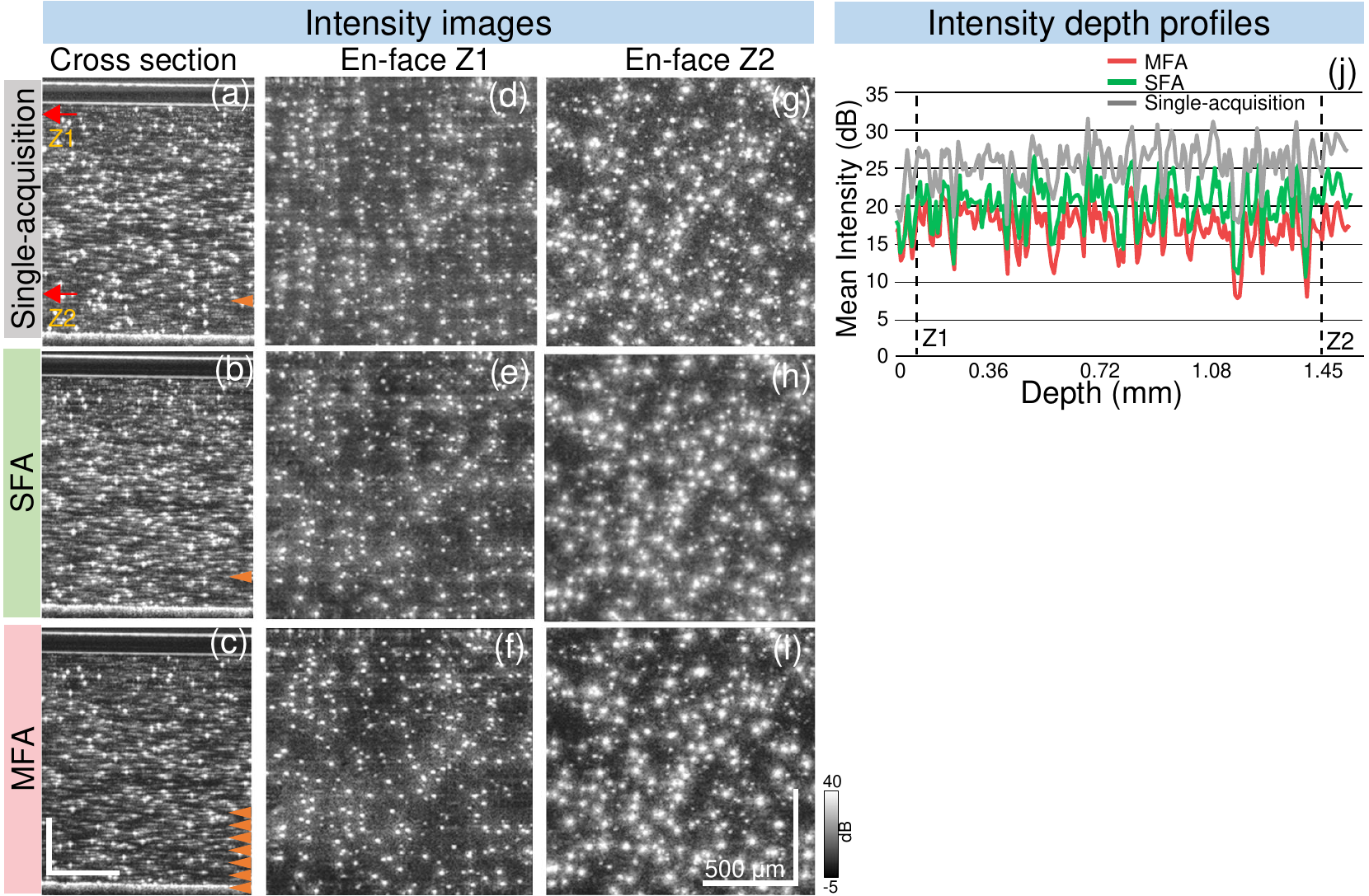}
	\caption{
		OCT intensity results for the scattering phantom. 
		(a)--(c) are the cross-sectional intensity images of the phantom. 
		(d)--(i) show the \enface intensity images at two depths (Z1 and Z2) indicated by the arrows in (a) and by the dashed lines in (j). 
		At depth Z1, compared with the single-acquisition results, both the SFA and MFA images show an intensity reduction in the background gel regions. 
		While at depth Z2, only the MFA image shows an obvious intensity reduction in the background. 
		The mean intensity depth profiles of the three methods in (j) show that both SFA and MFA profiles have lower mean intensity than the single-acquisition profile, while MFA shows more reduction than the SFA as the depth increases. 
		The focal positions are indicated by orange arrowheads in (a)--(c). 
	}
	\label{fig:phnatom_int}
\end{figure}

\begin{figure}
	\centering\includegraphics[width=13cm]{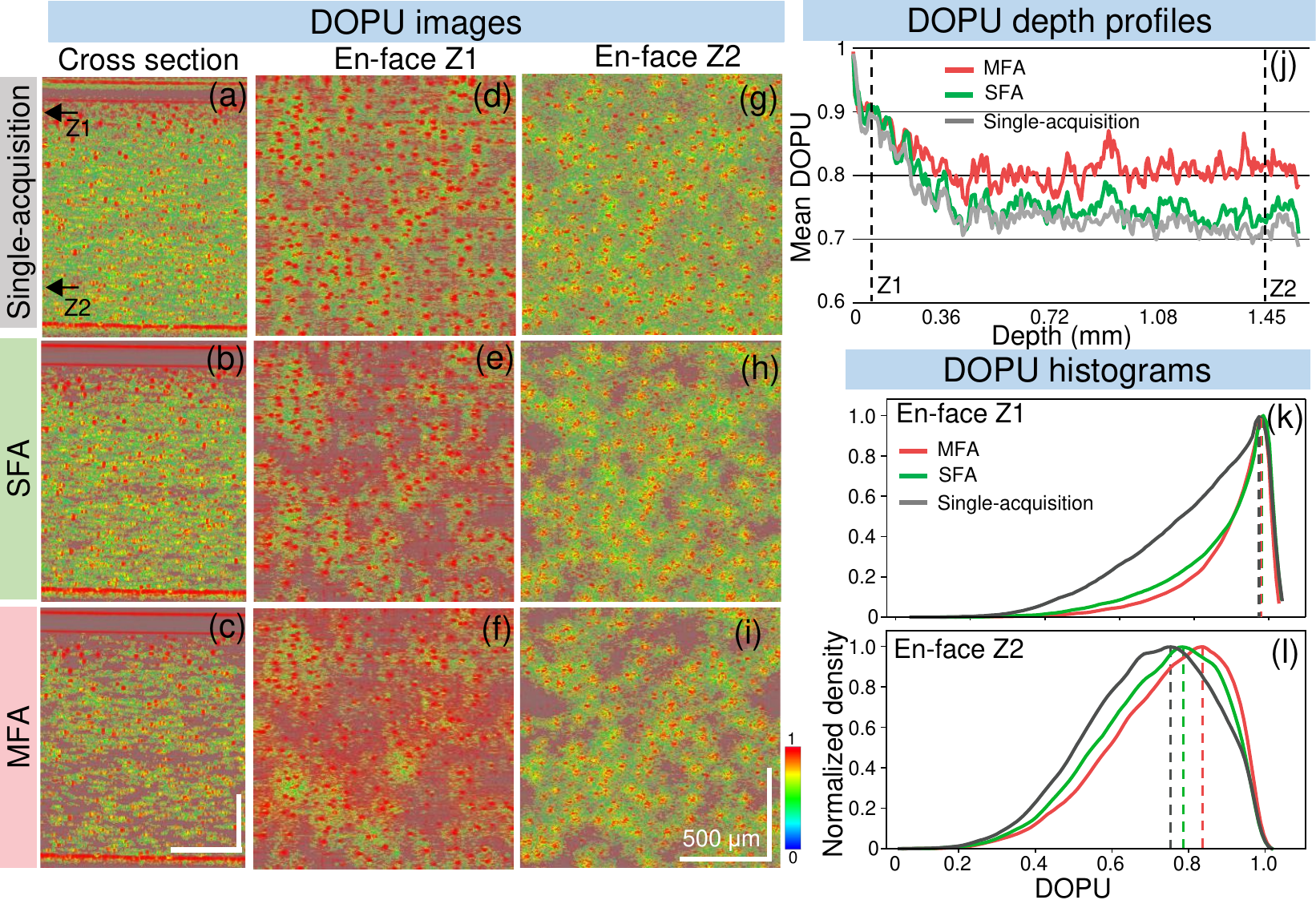}
	\caption{
		OCT intensity and DOPU measurement results for the scattering phantom.
		(a)--(c) show the cross-sectional DOPU images of the phantom, respectively. 
		(d)--(i) show the \enface DOPU images at two depths (Z1 and Z2) indicated by the arrows in (a) and by the dashed lines in (j).
		The polystyrene particles show high DOPU (red) and the surrounding space shows low DOPU (yellow).
		When compared with the single-acquisition results, the SFA and MFA images show a reduction in the low DOPU regions.
		The mean DOPU depth profiles of the three methods in (j) show that the DOPU of the MFA method is higher than that of the other methods.
		(k) and (l) are normalized histograms computed from (d)--(f) and (g)--(i), respectively. 
		The dashed vertical lines in (k) and (l) indicate the modes of each histogram. 
		The focal positions are indicated by orange arrowheads in (a)--(c). 
		The histogram at the superficial depth of Z1 in (k) shows the reduction in the low DOPU pixels in the SFA and MFA images when compared with the single-acquisition image. 
		The histogram at the deeper depth of Z2 in (l) demonstrates the increased mode values of the DOPU for the SFA and MFA when compared with the single-acquisition case.
	}
	\label{fig1:phnatom_dopu}
\end{figure}
The cross-sectional and \enface images of the intensity and DOPU of the phantom are shown in Fig.\@ \ref{fig:phnatom_int}(a)--(i) and Fig.\@ \ref{fig1:phnatom_dopu}(a--i), respectively, which are obtained using the single-acquisition, SFA, and MFA methods. 
The depth profiles of the mean intensity and mean DOPU obtained by these three methods are shown in Fig.\@ \ref{fig:phnatom_int}(j) and Fig.\@ \ref{fig1:phnatom_dopu}(j), respectively. 
These profiles were computed by averaging 10,000 A-lines (50 B-scans, with 200 A-lines per B-scan).
The normalized histograms of the DOPU shown in Figs.\@ \ref{fig1:phnatom_dopu}(k) and (l) were computed from the \enface DOPU images at two depths using a shifted average histogram method \cite{Freedman1981} with a bin size of 0.03125. 
%Previous reference {Anderson_2016}.
An intensity threshold for the noise floor of +5 dB was applied to exclude the pixels with very low scattering intensities when computing the mean DOPU depth profiles and the corresponding histograms. 

Among the three cross-sectional intensity images [Figs.\@ \ref{fig:phnatom_int}(a)–(c)], the MFA image shows lower background intensity in most of the regions of the scattering layer when compared with the other two images.
In the \enface images at a superficial depth Z1, both SFA and MFA images [Figs.\@ \ref{fig:phnatom_int}(e) and (f)] show reduced background intensity in the gel regions than the single-acquisition image [Fig.\@ \ref{fig:phnatom_int}(d)]. 
At a deeper depth Z2, only in the MFA image [Fig.\@ \ref{fig:phnatom_int}(i)] shows an obvious background reduction. 
The depth intensity profiles show that both SFA and MFA [Fig.\@ \ref{fig:phnatom_int}(j), green and red curves, respectively] show lower mean intensities than the single-acquisition (gray curve). 
The MFA shows higher mean-intensity reduction than the SFA at a deeper depth.

In addition, the imperfection of the mixture between the gel and microparticles may cause a spatially-variable distribution of scatter density, leading to a variable distribution of MS signals. 
Hence, the signal intensity in the gel regions may spatially vary within the scattering phantom. 
This effect can be mitigated by averaging multiple A-lines when plotting the depth profiles. 

In the cross-sectional single-acquisition DOPU image [Fig.\@ \ref{fig1:phnatom_dopu}(a)], the polystyrene microparticles show high DOPU (red), while the surrounding space shows low DOPU (yellow).
Because the ultrasound gel may not scramble the polarization, this appearance of low DOPU may be the artifact caused by the overlaying MS signals.
The mean DOPU depth profile [Fig.\@ \ref{fig1:phnatom_dopu}(j), gray curve] shows a sharp reduction along the depth until approximately 0.36 mm, and the profile becomes stably low in the deeper regions.  
When compared with the single-acquisition DOPU, the SFA cross-sectional DOPU image shows slightly fewer low-DOPU artifacts [Fig.\@ \ref{fig1:phnatom_dopu}(b)].
The mean-DOPU depth profile of the SFA image [Fig.\@ \ref{fig1:phnatom_dopu}(j), green curve] is slightly higher than that of the single-acquisition method, but the difference is not evident.
In contrast, the MFA image shows an evident reduction in the low-DOPU artifacts [Fig.\@ \ref{fig1:phnatom_dopu}(c)], and its mean-DOPU depth profile shows much higher mean DOPU values than those observed for the other two methods [Fig.\@ \ref{fig1:phnatom_dopu}(j)]. 

Figures \ref{fig1:phnatom_dopu}(d), (e), and (f) show the \enface DOPU images acquired at a superficial depth [Z1, as indicated in Fig.\@ \ref{fig1:phnatom_dopu}(a)].
In the single-acquisition image [Fig.\@ \ref{fig1:phnatom_dopu}(d)], the polystyrene particles show high DOPU and are enclosed by the low-DOPU artifact.
The low-DOPU artifact area is reduced in the SFA image [Fig.\@ \ref{fig1:phnatom_dopu}(e)] because of the reduction in the standard noise (i.e., shot noise, relative intensity noise, and detection noise) caused by the complex averaging process\cite{Szkulmowski_OE2013,Baumann_BOE2019}.
Further reduction in the low-DOPU artifact is observed in the MFA image [Fig.\@ \ref{fig1:phnatom_dopu}(f)].
Normalized histograms of these \enface images are shown in Fig.\@ \ref{fig1:phnatom_dopu}(k).
Although the histograms of the SFA and MFA images are sharper than that of the single-acquisition image, the mode values for these three images are all high and are nearly identical, with values of 0.967 (single-acquisition), 0.968 (SFA), and 0.968 (MFA).
These consistently high mode values indicate that there are fewer MS signals at this superficial depth, which would also explain the nearly identical histogram shapes for the SFA and MFA methods. 
In addition, the sharpening of the histogram may be accounted for by the reduction in the standard noise.

At the deeper depth of Z2 [as indicated in Fig.\@ \ref{fig1:phnatom_dopu}(a)], more low-DOPU artifacts are presented in the single-acquisition \enface image [Fig.\@ \ref{fig1:phnatom_dopu}(g)] than in the image at the superficial depth Z1.
The SFA image [Fig.\@ \ref{fig1:phnatom_dopu}(h)] shows a reduction in the area of the low-DOPU artifacts [Fig.\@ \ref{fig1:phnatom_dopu}(g)] and the MFA image [Fig.\@ \ref{fig1:phnatom_dopu}(i)] shows a further reduction in the area of these artifacts.
In the corresponding histograms [Fig.\@ \ref{fig1:phnatom_dopu}(l)], the mode value is low for the single-acquisition image (0.746), is higher for the SFA image (0.778), and is then at its highest for the MFA image (0.832).
The low mode value of the single-acquisition image may be caused by the standard noise and MS signals.
The higher mode value of the SFA image may be accounted for by the reduction in the standard noise, and the even higher mode value of the MFA image can be accounted for by the reduction in the MS signals.
These findings indicate that the MFA method is beneficial, particularly in the deeper regions.

\subsubsection{Birefringence of the phantom}  
\label{sec:result:phantom_biref}
\begin{figure}
	\centering\includegraphics[width=12cm]{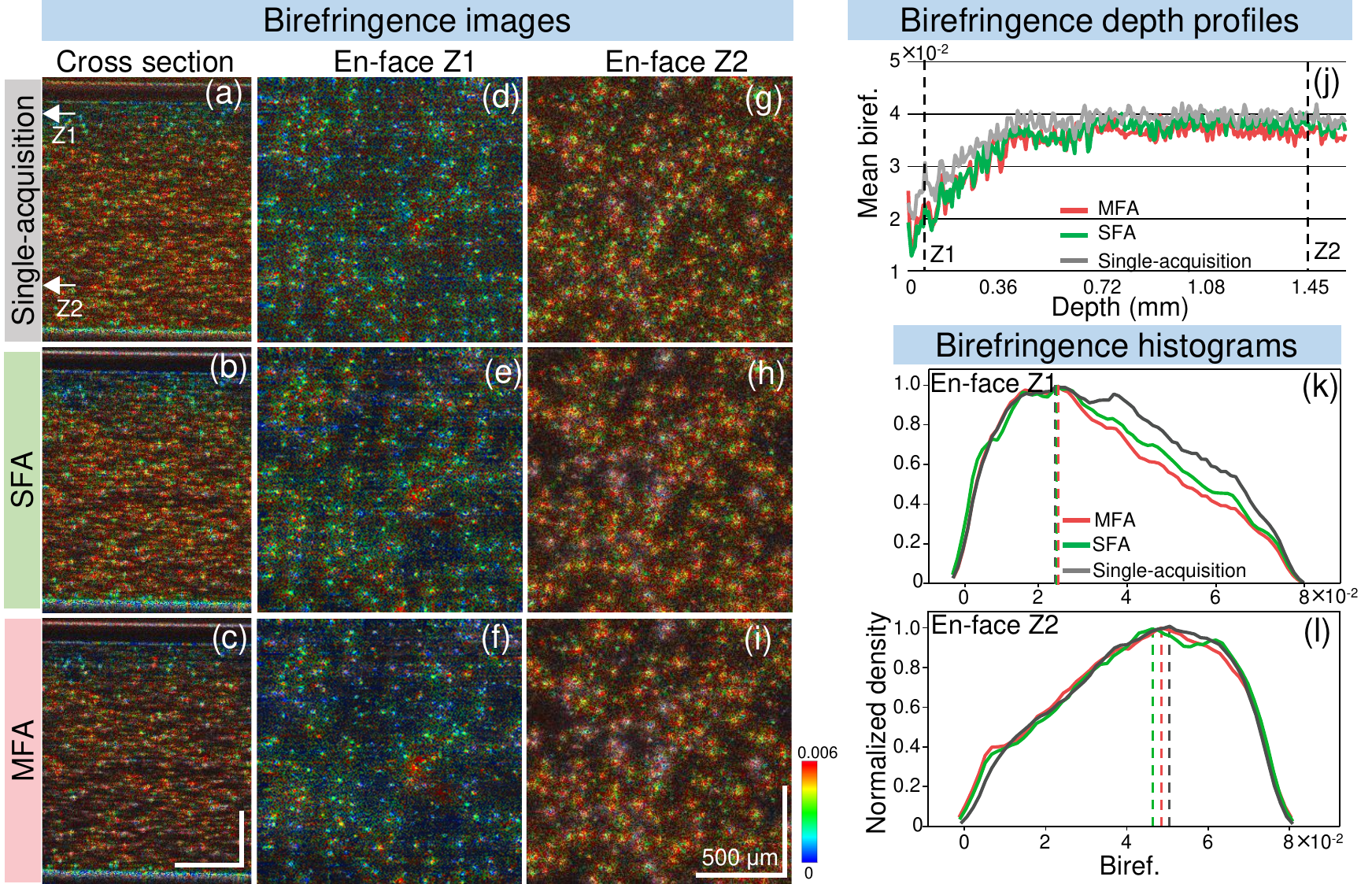}
	\caption{
		Birefringence measurement results for the scattering phantom.
		(a)--(c) and (d)--(i) show the cross-sectional and \enface slices of the birefringence images, respectively.
		These images correspond to those of Figs. \@ \ref{fig1:phnatom_dopu}(a)--(i). 
		At the superficial depth of Z1, the polystyrene particles show moderately high birefringence (green) and are surrounded by a moderately high birefringent halo. 
		The halo regions are reduced by the MFA method when compared with the other methods.
		At the deeper depth (Z2), the particles are surrounded by a halo with high birefringence, which may be artifacts caused by the dominance of the MS signals.
		(j) shows the mean birefringence depth profiles for the three methods.
		Here, the single-acquisition method shows slightly higher birefringence than the other methods.
		(k) and (l) show the histograms computed from (d)--(f) (at depth z1) and from (g)--(i) (at depth Z2), respectively. 
		The dashed vertical lines in (k) and (l) indicate the modes of each histogram. 
		The axial label ``biref.\@'' represents an abbreviation of birefringence.
		Although the histograms at the deeper depth in (l) do not show evident differences among the three methods, the corresponding histograms at the superficial depth in (k) showed a reduction in the high-birefringence pixels with the SFA and MFA methods. 
	}
	\label{fig2:phnatom_biref}
\end{figure}
Figures \ref{fig2:phnatom_biref}(a)--(i) show the birefringence images of the phantom, where the positions in the sample are identical to those in Fig.\@ \ref{fig1:phnatom_dopu}. 
At the superficial depth (Z1), the polystyrene particles that show moderate birefringence (green) are surrounded by a moderately birefringent (green) halo.
Because the surrounding gel was not birefringent, the moderately birefringent halo may be an artifact.
In the JM-OCT method, the birefringence can be artifactually high if the SNR is sufficiently low.
This artifactual elevation of the birefringence will be discussed in detail later in Section \ref{sec:discussion:birefElevation}.
When compared with the single-acquisition image [Fig.\@ \ref{fig2:phnatom_biref}(d)], the MFA image [Fig.\@ \ref{fig2:phnatom_biref}(f)] shows a less moderately birefringent (green) halo.
This difference may be caused by the reduction in the standard noise and the MS signals.

It should be noted that the high birefringence (green) layer at the bottom of the scattering phantom was the surface of the mending tape glued on a glass slip. 
Because of its manufacturing process, the mending tape is expected to have birefringence

In the deep region (Z2), both the particles and the surrounding gel showed high birefringence (indicated by colors ranging from yellow to red).
This may be caused by the predominance of the MS signals, and the MFA method could not show evident improvements for this high-scattering phantom.

Figure \ref{fig2:phnatom_biref}(j) shows the mean birefringence depth profiles, where the three profiles almost overlay each other, with the exception that the single-acquisition profile shows slightly higher birefringence values.
Figures \ref{fig2:phnatom_biref}(k) and (l) are the birefringence histograms obtained at depths Z1 (superficial) and Z2 (deep), respectively. 
At the superficial depth (Z1), some parts of the high birefringence appearance (i.e., the right shoulder of the histogram) are reduced by the SFA method and further reduced by the MFA method.
This reduction may occur because of the reduction in the standard noise (from single-acquisition to SFA) and the reduction in the MS signals (from SFA to MFA), and this behavior is consistent with the image appearances.
At the deeper depth (Z2), the histograms are all nearly identical to each other.
The ineffectuality of the SFA and MFA methods is also confirmed by the appearances of the images, and it may be caused by the scattering being too strong.

\subsection{Postmortem medaka fish}  
\label{sec:result:fish}
First, we compare the OCT intensity, DOPU, and birefringence images of the postmortem medaka fish acquired using the three methods objectively.

In addition, for quantitative comparison of the measured DOPU and birefringence, we manually selected four ROIs within the cross-sectional images.
The size of each ROI is 15 (axial) $\times$ 35 (lateral) pixels (0.2 mm axial $\times$ 0.1 mm lateral).
The ROIs are indicated by white boxes in both Figs.\@ \ref{fig4:fish_dopu}(a) and \ref{fig5:fish_biref}(a). 
We selected the skin and muscle regions in the upper and lower halves of the fish, where the terms upper and lower refer to the upper and lower areas of the image, i.e., they correspond to the right and left sides of the fish.
Because the fish body under investigation is symmetrical in terms of the right and left sides, the upper and lower ROIs should have similar DOPU and birefringence values as long as polarization artifacts do not exist.
We applied an intensity threshold of +5 dB to exclude pixels with very low scattering intensities from the computation of the statistics, i.e., the means and percentiles.

\subsubsection{OCT intensity images of the medaka fish} 
\label{sec:result:fish_int}
\begin{figure}
	\centering \includegraphics[width=9cm]{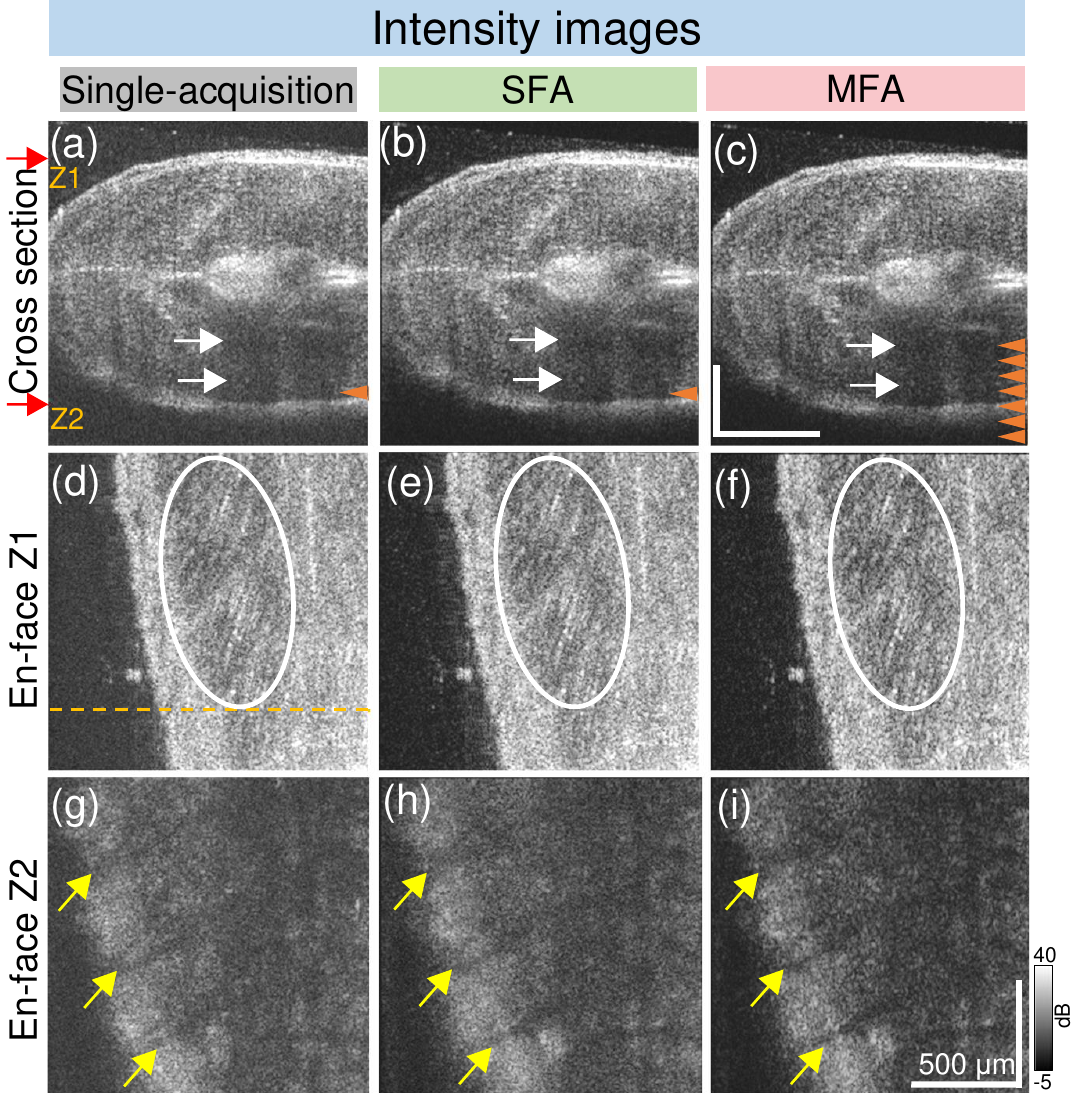}
	\caption{
		OCT intensity images of the postmortem medaka fish.
		(a)--(c) show the cross-sectional images and (d)--(i) show the \enface intensity images. 
		The white arrows in (a)--(c) indicate the shadows in the deeper region of the sample, which become darker in the MFA image than in the other images because of the reduction in the MS signals.
		The white ellipses in (d)--(f) indicate the fibrous muscle tissue.
		The contrast in this region is improved using the MFA method.
		The yellow arrows in (g)--(i) indicate low-intensity shadows, and the contrasts of these shadows are found to be improved by the MFA method.
		The red arrows in (a) and the orange dashed line in (d) indicate the locations in which the \enface and cross-sectional images were acquired, respectively. 
		The focal positions are indicated by the orange arrowheads shown in (a)--(c). 
	}
	\label{fig3:fish_int}
\end{figure}
Figure \ref{fig3:fish_int} shows the intensity images of the medaka fish for the single-acquisition, SFA, and MFA methods.
In the cross-sectional images [Fig.\@ \ref{fig3:fish_int}(a)--(c)], the shadow region on the lower side of the fish (white arrows) becomes darker in the SFA image than in the single-acquisition image, and is even darker in the MFA image.
This change may be caused by the reduction in the standard noise (from single-acquisition to SFA) and by the reduction in the MS signals (from SFA to MFA).

In the \enface images at the superficial depth (Z1) [Figs.\@ \ref{fig3:fish_int}(d)--(f)], an area that shows fibrous structures can be observed (indicated by white ellipses), and this is believed to be the muscle region. 
The contrast in this region is improved using the MFA method.

At the deeper depth (Z2), the improvement in the contrast is more evident [Figs.\@ \ref{fig3:fish_int}(g)--(i)].
The SFA image shows a slightly enhanced contrast between the tissues and some low-intensity structures (indicated by the yellow arrows), and the MFA image shows further enhancement. 
These low-intensity structures are the projection shadows that resulted from hyper scattering or from the absorption of the tissues at the shallower depths.

\subsubsection{DOPU of the medaka fish} 
\label{sec:result:fish_dopu}
\begin{figure}
	\centering \includegraphics[width=13cm]{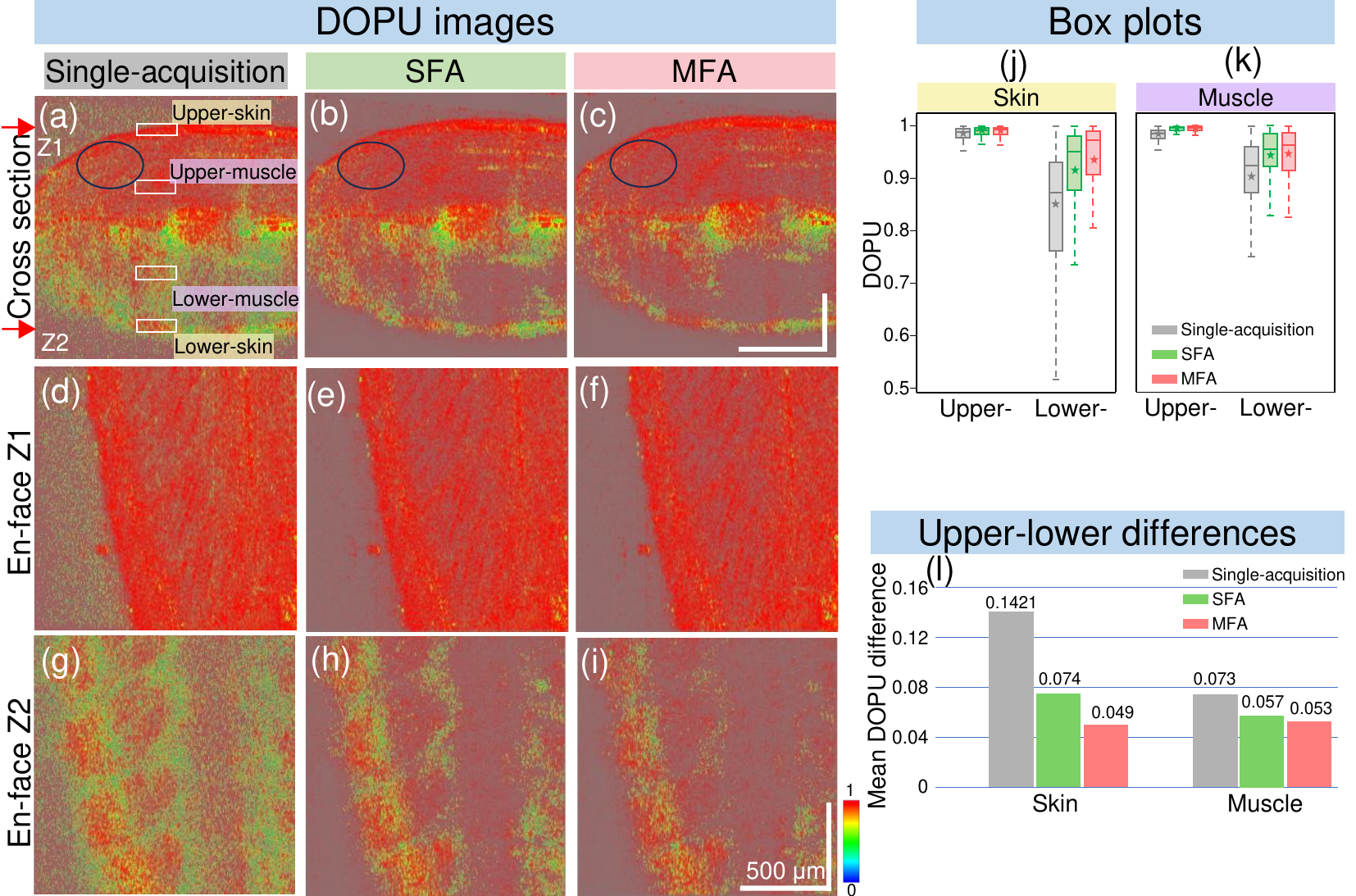}
	\caption{
		DOPU measurement results for the medaka fish.
		(a)--(c) show the cross-sectional images and (d)--(i) show the \enface DOPU images at two depths. 
		When compared with the single-acquisition and SFA images, the MFA images [(c) and (i)] show fewer low-DOPU artifacts in the deeper regions. 
		The white boxes in (a) indicate the ROIs selected for the quantitative analysis. 
		The black circles in (a--c) indicate some small low-DOPU dots which may be resulted from the standard noise. 
		(j) and (k) show the box plots of the DOPU values in the skin and muscle ROIs, respectively. 
		In each box, the upper and lower boundaries denote the upper and lower quartiles, respectively, the center lines denote the medians, the stars denote the means, and the whiskers denote the minimum and maximum values. 
		In the deeper region (i.e., the area with the lower ROIs), the MFA produces increased DOPU values, which may indicate the reduction of the MS signals caused by the MFA method.
		(l) shows the differences in the mean DOPU between the lower and upper ROIs, which is called the upper-lower difference, for the skin and muscle regions. 
		Both the SFA and MFA methods provided smaller upper-lower differences of DOPU than the single acquisition method.
	}
	\label{fig4:fish_dopu}
\end{figure}
Figure \ref{fig4:fish_dopu} shows the DOPU images of the medaka fish. 
In the single-acquisition images, both the skin and muscle regions in the upper half of the fish show homogeneously high DOPU [see the upper half of Fig.\@ \ref{fig4:fish_dopu}(a) (cross-section) and Fig.\@ \ref{fig4:fish_dopu}(d) (\enface at superficial depth Z1)].
We also see a few small low-DOPU (yellow) dots, such as being found in the black circles in Fig.\@ \ref{fig4:fish_dopu}(a--c), which may represent the artifacts caused by the standard noise.
These small dots are eliminated by both the SFA and MFA methods.

The lower half of the fish shows numerous low-DOPU signals [see the lower half of Fig.\@ \ref{fig4:fish_dopu}(a) (cross-section) and Fig.\@ \ref{fig4:fish_dopu}(g) (\enface at the deeper depth Z2)]. 
On the basis of the symmetry of the fish body, these low-DOPU signals may be artifacts.
When compared with the single-acquisition images, the SFA images show reduced area of artifacts [Figs.\@ \ref{fig4:fish_dopu}(b) and (h)], and the MFA images show a further reduction of the area [Figs.\@ \ref{fig4:fish_dopu}(c) and (i)]. 

Comparing the DOPU images with the intensity images in Fig.\@ \ref{fig3:fish_int}, most of the regions in the upper half of the fish exhibit both high intensity and high DOPU. It can be explained by the dominance of the SS signals in the upper half of the fish. 
On the other hand, as the depth goes deeper than around the middle of the fish, many low-DOPU artifacts are presented, and the signal intensity is slightly attenuated as the depth increases. Several regions that exhibit moderate intensity and low DOPU can be noticed. The sudden change of the DOPU in the lower half might be due to some highly scattering anatomic structures around the dorsal midline of the fish, such as bones or connective tissues. This phenomenon has also been revealed in several other researches\cite{Lichtenegger_BOE2022,Lichtenegger_SciRep2022}.

The box-whisker plots shown in Figs.\@ \ref{fig4:fish_dopu}(j) and (k) compare the results for the three methods for the upper and lower ROIs for the skin [Fig.\@ \ref{fig4:fish_dopu}(j)] and for the muscle [Fig.\@ \ref{fig4:fish_dopu}(k)] ROIs.
Here, the boxes indicate the upper and lower quartiles, the center bar denotes the median, and the whiskers denote the maximum and minimum values.
At the upper ROIs, for both the skin and the muscle, all three methods produced similarly high DOPU because the MS signals are not dominant at these superficial depths.
Although the mean DOPU values obtained with the single-acquisition method are close to 1, they are improved (i.e., they become higher) by the SFA method, and are improved further by the MFA method.

At the lower ROIs, these improvements are more evident.
For both the skin and muscle ROIs, the single-acquisition method shows artifactually low mean DOPU values, and this artifactual reduction tendency is clearly mitigated by the SFA method. 
For the skin ROI, the mean DOPU was increased further (i.e., improved) by the MFA method.
For the muscle ROI, the mean DOPU was increased further by the MFA method (from 0.93 for the SFA to 0.94 for the MFA).

The differences between the mean DOPU values for the upper and lower ROIs (i.e., the values obtained by subtracting the lower ROI mean from the upper ROI mean) are depicted in Fig.\@ \ref{fig4:fish_dopu}(l).
Because of the symmetry of the fish’s anatomy, a smaller difference represents a better reduction in the artifact.
When compared with the other two methods, the MFA method provided the smallest upper-lower difference values for both the skin and muscle ROIs. 

\subsubsection{Birefringence of the medaka fish} 
\begin{figure}
	\centering\includegraphics[width=13cm]{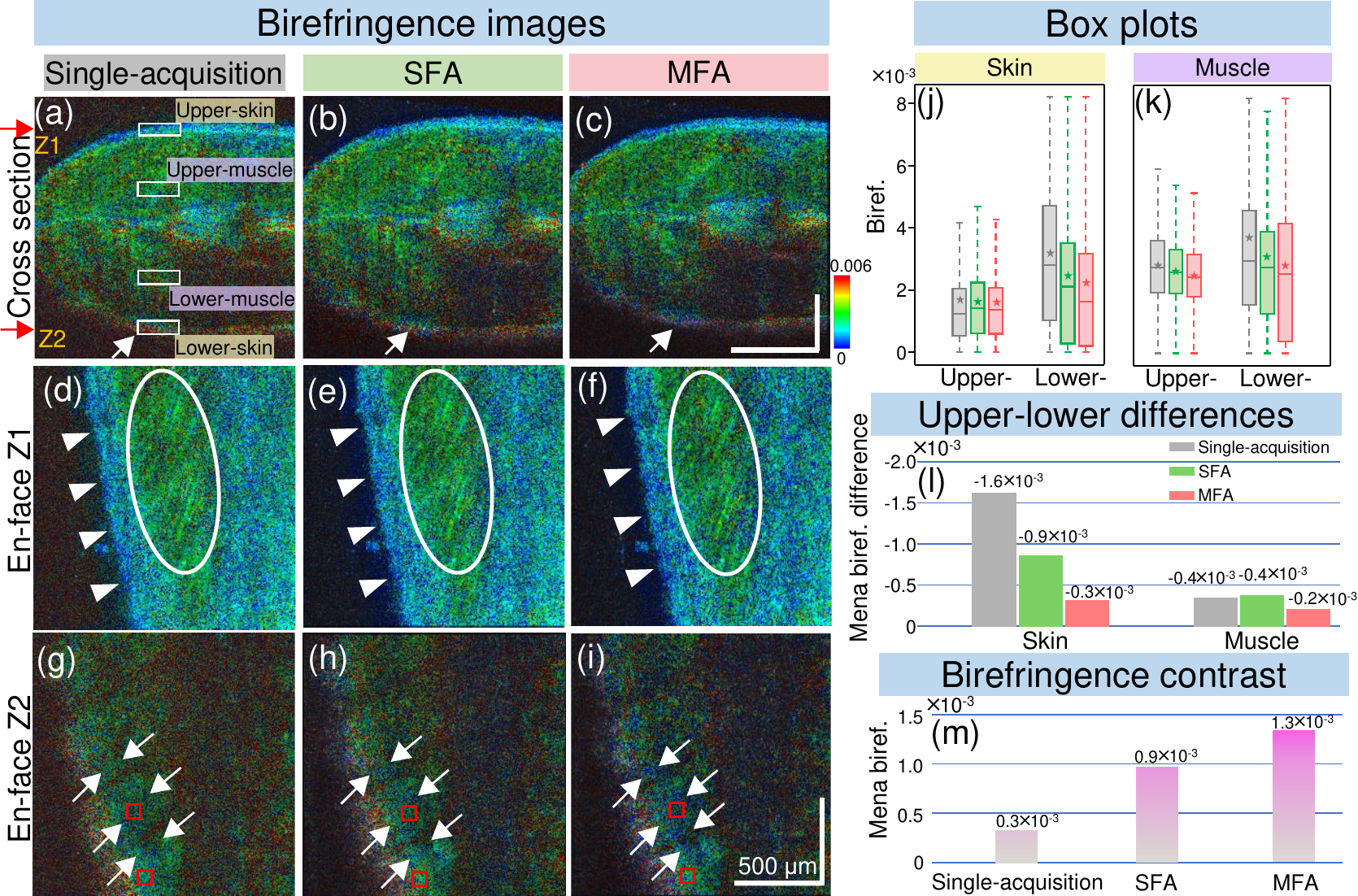}
	\caption{
		Birefringence imaging results of the medaka fish.
		(a)--(c) show the cross-sectional birefringence images and (d)--(i) show the \enface birefringence images at two depths (the superficial depth Z1 and the deeper depth Z2).
		At depth Z1 in (d)--(f), the \enface images of the three methods show a muscle region (indicated by white ellipses) that exhibits high birefringence (green).
		In the single-acquisition image in (d), the neighboring skin region (indicated by the arrowheads) shows moderate birefringence (blue to green), but this birefringence is slightly reduced by the SFA method in (e) and the MFA method in (f).
		At depth Z2 in (g)--(i), only the MFA image shows a region with low birefringence (indicated by the white arrows), which is the skin in the lower half of the fish [see also the region indicated by the white arrow in (c)]. 
		(j) and (k) show the box plots of the birefringence values in the four ROIs that are indicated in (a). 
		(l) shows the differences in the mean birefringence between the lower and upper ROIs for both the skin and muscle regions. 
		The SFA and MFA methods reduce the birefringence, especially in the lower ROIs.
		(m) shows the ``birefringence contrasts'' of the three methods, where the birefringence contrast is defined as the difference between the mean birefringence values in the low- and high-birefringence regions [red boxes in (g)--(i)]. 
		The contrast is improved by the SFA method and is improved further by the MFA method.
		The axial label ``biref.\@'' is an abbreviation of birefringence.
	}
	\label{fig5:fish_biref}
\end{figure}
Figure \ref{fig5:fish_biref} presents the birefringence images of the medaka fish.
In the cross-sectional images [Figs.\@ \ref{fig5:fish_biref}(a)--(c)], the skin in the upper half of the fish shows low birefringence (blue) for all three methods. 
In contrast, the skin in the lower half shows high birefringence (red to green) in the single-acquisition and SFA images [arrows in Figs.\@ \ref{fig5:fish_biref}(a) and (b)], but the MFA image shows lower birefringence (blue) than the other methods [arrow in Fig.\@ \ref{fig5:fish_biref}(c)]
This reduction in the birefringence also can be observed in the \enface images at the deeper depth (Z2) [arrows in Figs.\@ \ref{fig5:fish_biref}(g)--(i)].
The \enface images at the superficial depth (Z1) [Figs.\@ \ref{fig5:fish_biref}(d)--(f)] also show slight reductions in the birefringence (indicated by the white arrowheads) produced by the SFA and MFA methods. 
This reduction may be less obvious because the MS signals were less.

Two regions showing high and low birefringence were selected using a 15 $\times$ 15 pixel window in the deeper-depth \enface images [red boxes in Figs.\@ \ref{fig5:fish_biref}(g)--(i)].
The difference between the mean birefringence values of these two regions, which was computed by subtracting the mean of the low birefringence region from that of the high birefringence region and denotes the birefringence contrast, was plotted in Fig. \@\ref{fig5:fish_biref}(m).
When compared with the single-acquisition method, the SFA method shows a higher birefringence contrast, and the MFA method shows an even higher birefringence contrast.
This finding highlights the ability of the MFA method to enhance the birefringence contrast between the skin and the neighboring muscle tissues in the deeper regions.

Figures \ref{fig5:fish_biref}(j) and (k) show box-whisker plots for the skin ROIs [Fig.\@ \ref{fig5:fish_biref}(j)] and the muscle ROIs [Fig.\@ \ref{fig5:fish_biref}(k)]. 
These ROIs are indicated in Fig.\@ \ref{fig5:fish_biref}(a).
At the upper ROIs of both the skin and the muscle, all methods show similarly low birefringence.
In contrast, the SFA method shows reduced mean birefringence in the lower ROIs when compared with the single-acquisition method, and this birefringence is reduced further by the MFA method.

Because of the anatomical symmetry of the fish, the mean birefringence should be similar for the upper and lower ROIs.
The plots of the upper-lower differences [Fig.\@ \ref{fig5:fish_biref}(l)] demonstrate that the MFA method achieves the closest similarity in these mean birefringence values for both the skin and the muscle ROIs among the three methods.

In summary, the polarization metrics (i.e., the DOPU and the birefringence) are found to be altered by the MS signals in the deeper regions. 
The results indicate that the MFA method is less susceptible to these alterations than both the single-acquisition and SFA methods. 
In other words, the MFA method reduces these artifactual alterations effectively. 
A discussion of the interaction between the MS signal and the estimates of the DOPU and birefringence will be presented in Sections \ref{sec:discuss:interpretation} and \ref{sec:discuss:MSimpact}.

\section{Discussions}
\subsection{Interpretation of polarization artifacts associated with multiple scattering}
\label{sec:discuss:interpretation}
In this work, the DOPU and birefringence measured by PS-OCT were found to be prone to the MS signals in the deeper regions.
Adie \etal found that the presence of MS signals increases the randomness of the measured polarization states\cite{adie_OEO_2007}.
This randomization can then lead to low DOPU estimation.
Specifically, as the depth in the sample increases, stronger MS signals will be present, and these signals cause the erroneously low DOPU.

Understanding of the interaction between the MS signal and the birefringence measurement remains an open issue. 
In this study, we applied a MAP estimator to perform the maximum likelihood estimation of the birefringence. 
However, this method did not account for the presence of the MS signal and it was thus not expected to yield accurate birefringence estimation in such cases. 
Nonetheless, we observed a reduction in the skin birefringence in the deeper regions of the medaka fish after applying the MFA method, which was consistent with the appearance of the skin birefringence in the superficial regions. 
This finding indicates that removal of the MS signal can improve the birefringence estimation accuracy. 

Preliminary results about the mathematical formulations of the MS-effects on the polarization measurements are presented in the following sections (Sections \ref{sec:MsInBiref} and \ref{sec:MsDOPU}).

\subsection{Mathematical description of MS-signal's impact on birefringence and DOPU measurements}
\label{sec:discuss:MSimpact}
\subsubsection{Impact on birefringence measurement}
\label{sec:MsInBiref}
In our JM-OCT method, the birefringence is defined as the local phase retardation. 
The local phase retardation is the phase retardation of a measured local Jones matrix, which can be expressed as follows, 
\begin{equation}	
	\label{eq:ljm}
	\bJ_m (z_2) \bJ_m^{-1} (z_1) = \Jout \bJ_s(z_2) \bJ_s^{-1}(z_1) \Jout^{-1} . 
\end{equation}
Here, $\bJ_m(z)$ represents the measured Jones matrix at a depth $z$, and the lateral positions $x$ and $y$ are omitted for simplicity.
$z_1$ and $z_2$ are two close depths.
The local phase retardation is the phase retardation between these two depths \cite{Yasuno_IEEE2023}. 
$\Jout$ is the Jones matrix of the collection optics of the system.
The defocus and the lateral positions $x$ and $y$ are omitted for simplicity, and it is assumed that there is no MS signal.

Note that $\bJ_s(z_2) \bJ_s^{-1}(z_1)$ is a round-trip Jones matrix of a sample between the two depths.
Since this round-trip Jones matrix is a matrix similar to the local Jones matrix $\bJ_m(z_2) \bJ_m^{-1}(z_1)$, the eigenvalues of the local Jones matrix are identical to those of $\bJ_s(z_2) \bJ_s^{-1}(z_1)$.
Hence, the phase retardation of the local Jones matrix [Eq.\@ (\ref{eq:ljm})] is identical to that of the round-trip Jones matrix between $z_1$ and $z_2$.

By considering the MS components, the measured Jones matrix can be written as
\begin{equation}
	\label{eq:jm_ssms}
	\bJ_m' (z) = \bJ_m^{SS} (z) + \bJ_m^{MS} (z),
\end{equation}
where $\bJ_m^{SS}$ and $\bJ_m^{MS}$ are the SS and MS components of the measured Jones matrix and are formed by $S^j_{SS}$ and $S^j_{MS}$ of Section \ref{sec:principle:theory}, respectively. 
By substituting $\bJ_m'$ into $\bJ_m$ of Eq.\@ (\ref{eq:ljm}), the local Jones matrix affected by the MS signal can be derived as
\begin{equation}
	\label{eq:ljm_ssms}
	\bJ_m' (z_2) \bJ_m'^{-1} (z_1) = 
	\left[
		\bJ_{m}^{SS} (z_2) + \bJ_{m}^{MS} (z_2)
	\right]
	\left[
		\bJ_{m}^{SS} (z_1) + \bJ_{m}^{MS} (z_1)
	\right]^{-1}.
\end{equation}

\newcommand{\mA}{{\mathbf{A}\xspace}}
\newcommand{\mB}{{\mathbf{B}\xspace}}
\newcommand{\mI}{{\mathbf{I}\xspace}}
\newcommand{\inv}{{^{-1}\xspace}}
By using a general relationship of two matrices $\mA$ and $\mB$, $(\mA + \mB)\inv = \mA\inv - (\mI + \mA\inv \mB)\inv \mA\inv \mB \mA\inv$ (see Supplementary 1 for proof), Eq.\@ (\ref{eq:ljm_ssms}) becomes
\begin{equation}
\label{eq:ljm_ssms_d}
\begin{split}
	\bJ_m' (z_2) \bJ_m'^{-1} (z_1) =&
	\left[
	\bJ_{m}^{SS} (z_2) + \bJ_{m}^{MS} (z_2)
	\right] \\
	&\left[
	\bJ_{m}^{SS} (z_1)\inv - 
	\left(
	\mI + \bJ_{m}^{SS} (z_1)\inv \bJ_{m}^{MS} (z_1)
	\right)\inv
	\bJ_{m}^{SS} (z_1)\inv
	\bJ_{m}^{MS} (z_1)
	\bJ_{m}^{SS} (z_1)\inv
	\right]\\
	=&\quad \bJ_{m}^{SS} (z_2) \bJ_{m}^{SS} (z_1)\inv\\
	& + \bJ_{m}^{MS} (z_2) \bJ_{m}^{SS} (z_1)\inv\\
	& - \bJ_{m}^{SS} (z_2)\left(
		\mI + \bJ_{m}^{SS} (z_1)\inv \bJ_{m}^{MS} (z_1)
		\right)\inv
		\bJ_{m}^{SS} (z_1)\inv
		\bJ_{m}^{MS} (z_1)
		\bJ_{m}^{SS} (z_1)\inv\\
	& - \bJ_{m}^{MS} (z_2)\left(
		\mI + \bJ_{m}^{SS} (z_1)\inv \bJ_{m}^{MS} (z_1)
		\right)\inv
		\bJ_{m}^{SS} (z_1)\inv
		\bJ_{m}^{MS} (z_1)
		\bJ_{m}^{SS} (z_1)\inv.	
\end{split}
\end{equation}
This equation is the formulation of the local Jones matrix obtained when MS signals exist.
The first term in the last part of the equation is identical to the ideal local Jones matrix that is composed only of SS signals, while the other three terms originate from the interactions between SS and MS signals, causing the birefringence artifacts.
It should be noted that the birefringence (i.e., the local phase retardation) is obtained through a nonlinear computation from Eq.\@ (\ref{eq:ljm_ssms_d}).
Hence, the artifactual birefringence would not be additive, even though the SS-MS interaction terms are additive.

\subsubsection{Impact on DOPU measurement}
\label{sec:MsDOPU}
As discussed in the previous sections, the measured Jones matrix is the summation of the SS and MS components, as shown in Eq.\@ (\ref{eq:jm_ssms}). 
The MS signals can be considered as an additive noise and may artifactually reduce the DOPU value. 
Similar effects from standard (i.e., random) OCT noises have been described in Ref.\@ \cite{Makita_OL2014}.

Although we have used Makita's noise correction\cite{Makita_OL2014} to compute DOPU, this noise correction cannot compensate the effect of MS signals for two reasons.
At first, Makita's correction relies on predefined (i.e., measured) noise energies, typically obtained in a calibration measurement by blocking the probe beam, i.e., by isolating only the noise component.
On the other hand, the MS signals cannot be isolated in the calibration measurement, hence Makita's correction cannot correct the MS effect.
Second, Makita's correction assumes the noise is fully random, whereas the MS signals are not necessarily fully random.
This can be the other cause of the incompatibility of Makita's method to correct the DOPU artifact from the MS signals.

\subsection{Artifactual elevation of birefringence in the low signal-to-noise region}
\label{sec:discussion:birefElevation}
Although the SFA method cannot reduce the MS signals, the results for both the phantom and the fish showed that the SFA method reduces the measured birefringence values.
This can be explained by the noise susceptibility of the phase retardation and birefringence measurements of JM-OCT \cite{Makita2010OpEx, Duan2011OpEx}.
The phase retardation when measured by JM-OCT converges at approximately 2/3 $pi$ rad as the effective SNR decreases.
If the true phase retardation is low, this effect results in an erroneous up-shift in the measured phase retardation.
Because the birefringence used in this study is defined as a local phase retardation, the birefringence also suffered from this systemic error.
Although the MAP birefringence estimator used in this study can mitigate this effect, it remains effectual at very low SNR \cite{Kasaragod_OE2014, Kasaragod_BOE2017}.

This artifactual elevation of the birefringence can explain the reduction in the measured birefringence realized by the SFA method, because the SFA method increases the SNR by complex averaging.

\subsection{Limitations of the MFA implementation} 
\label{sec:discuss:limitation}
The relationship between the MS behavior and the polarization artifacts is not well understood at present.
Theoretical modeling and/or numerical simulations of the PS-OCT signal in an MS regime may provide a more comprehensive understanding and further insights in future work.

%%%
The quantification of the accuracy improvement achieved by MFA is not straightforward and is yet difficult. 
There are several challenges that complicate this quantification. 
One is the sample- and structure-dependency. 
The contribution of MS components highly varies among samples, i.e., the MS signals vary across different samples (sample-dependent variability) and within the structures of each individual sample (structure-dependent variability). 
This makes it challenging to standardize a quantification of improvement across different experimental conditions. 
The second issue is the lack of established quantification methods. 
Currently, there is no widely accepted methodology for quantifying the MS components not only for PS-OCT but also for intensity OCT imaging. 
This absence of a standardized approach brings a great challenge to precisely measuring the improvement by our MFA method. 
Third issue is the lack of ground truth images. 
To accurately quantify the improvement achieved by MFA, an exact ground truth of MS-free images is required. 
However, obtaining such data is difficult due to the complex nature of optical scatterings.
This difficulty makes the accurate evaluation of MFA impracticle.

As acknowledged in Section 4.2 in our previous work \cite{LZhu2023BOE}, the enhancement provided by MFA is limited. 
There is still a possibility to further enhance the MS reduction capability of MFA. 
A potential approach might be increasing the complexity of the wavefront pattern controllable by the tunable lens.
Lima \etal demonstrated a combination of a tunable lens and an electrode array\cite{Lima_OE2017}. 
This combination serves as a wavefront manipulation device capable of generating several complex low-order aberrations, enabling finer manipulations of the wavefront, and hence, higher decorrelation of MS signals.

Another limitation of the MFA method is the difficulty of performing \invivo imaging. 
Because the MFA method requires multiple volumetric imaging, it is vulnerable to the sample’s motion.
In other words, the sample should remain stationary during all the volumetric measurements, which have taken approximately 20 s to complete in the present study.

One possible solution is to integrate the MFA method with a high-speed PS-OCT system \cite{SouthBOE2014, HuberBOE2017, Kolb_Plos2019, Auksorius_BOE2020, Siyu_SciRep_2022}. 
Another potential solution is to use B-scan-based focus modulation, in which the depth position of the focus changes B-scan by B-scan, which means that the sample only need to remain stationary for a few B-scan acquisitions and not for a few volume acquisitions.
In our preliminary study, a set of B-scans with focus modulation was acquired in 86.2 ms \cite{Yiqiang_bios2023} using the same JM-OCT system.

Either of these two solutions, or a combination of them, may help achieve MS signals reduction for \invivo imaging and enhance visualization of the polarization features in deep tissues.

\subsection{Reproducibility of MS reduction ability of MFA}
\label{sec:discuss:reproducibility}
\begin{figure}
	\centering\includegraphics[width=10cm]{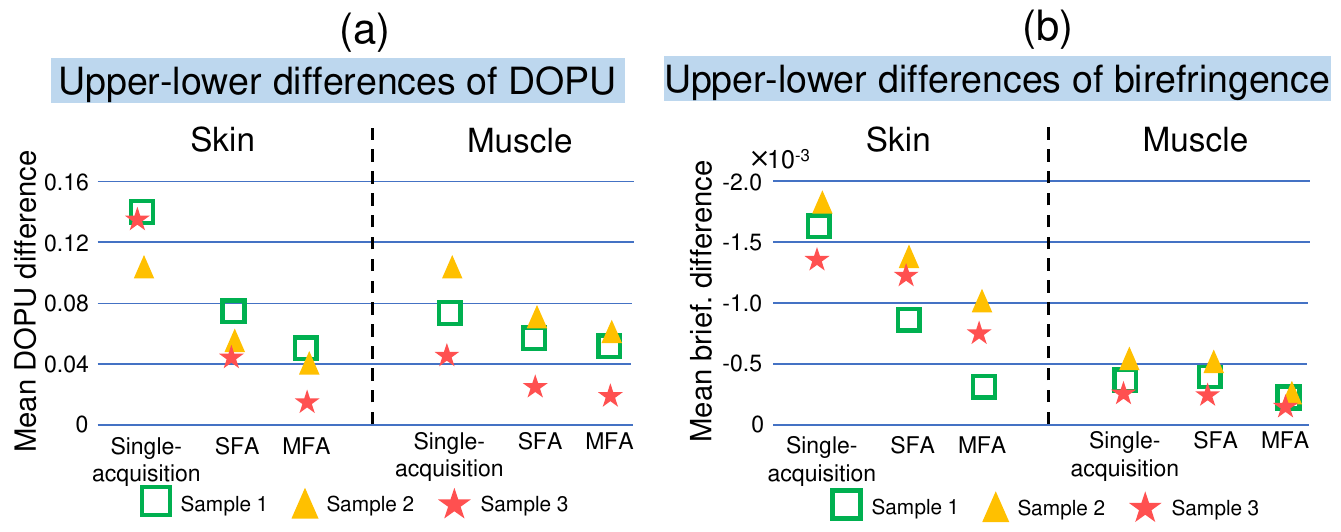}
	\caption{
		Upper-lower differences of the (a) DOPU and (b) birefringence results from three additional fishes. 
		Samples 2 and 3 were two adult medaka fishes, sample 1 is from the identical fish data that presented in the Result section. 
		More reduction in upper-lower differences by MFA than the other two methods may indicate a better MS reduction. 
		In the skin regions, MFA shows obvious reductions in the upper-lower differences of DOPU and birefringence than the other methods for all three samples. 
		In the muscle regions, MFA only gives moderate reduction than the SFA method. 
	}
	\label{fig:reproduce}
\end{figure}
To validate the reproducibility of the MFA method's ability to reduce the MS signals, we extended our experiment by measuring two additional postmortem adult medaka fishes which were approximately 6-months old. 
These new samples are referred to as sample 2 and 3, while the sample presented in the Result Section is referred to as sample 1. 
Samples 2 and 3 were measured with the same protocol to that of sample 1 (see Section \ref{sec:validation:protocol}).

Figure \ref{fig:reproduce} shows the upper-lower differences of mean DOPU and mean birefringence of the three samples. 
The results of sample 1 are identical to that presented in Figs.\@ \ref{fig4:fish_dopu} and \ref{fig5:fish_biref}. 
In the skin regions [left sides of (a) and (b)], all the three samples show consistent result where MFA demonstrates an obvious and higher reduction in the upper-lower differences than the other two methods. 
Additionally, in the muscle region, the MFA consistently shows the smallest upper-lower difference among the three methods. 
These consistent results suggest reasonable reproducibility of MFA performance.

The DOPU and birefringence images of samples 2 and 3 are presented in a supplementary material (Fig.\@ S1).

\subsection{Application in spectral-domain-OCT}
\label{sec:discuss:sdoct}
So far, the MFA method has been demonstrated only with a swept-source (SS)-OCT system. 
However, we believe there are no fundamental limitations preventing its application to spectral-domain (SD)-OCT. 
In SD-OCT, signal roll-off is typically more rapid than in SS-OCT, due to the relatively lower wavelength resolution of spectrometers compared to the instantaneous spectral line width of the light source in SS-OCT. 
While this higher signal roll-off in SD-OCT may result in lower SNR, it does not affect the ratio between SS and MS signals. 
Furthermore, the decorrelation of MS signals by the focus modulation is not affected regardless of whether SS-OCT or SD-OCT is chosen. 
Therefore, we believe that the MFA method to be applicable to SD-OCT as well.

\section{Conclusion}
In this work, we demonstrate a technique called MFA to mitigate the MS signals in Jones matrix PS-OCT. 
The proposed method was validated using a scattering phantom and a postmortem medaka fish, and it was demonstrated that the MFA method reduces the artifacts for both DOPU and birefringence effectively.
As a result, improved DOPU and birefringence image contrasts were obtained.
The proposed method is a cost-effective solution to improve the accuracy of quantitative polarization measurements. 
This work may have important implications for polarization-sensitive imaging and polarization properties quantification of thick biological tissues.

\section* {Acknowledgments}
Lida Zhu is supported by the China Scholarship Council through the Chinese Government Graduate Student Overseas Study Program.

\section*{Funding}
Core Research for Evolutional Science and Technology (JPMJCR2105); 
Japan Society for the Promotion of Science (21H01836, 22K04962); 
China Scholarship Council (201908130130).

% \disclosures 
\section*{Disclosures}
L. Zhu, S. Makita, Y. Lim, Y. Zhu, Y. Yasuno: Yokogawa Electric Corp. (F), Sky Technology (F), Nikon (F), Kao Corp. (F), Topcon (F). 
J. Tamaoki, M. Kobayashi: None.

\section* {Data, Materials, and Code Availability} 
Data underlying the results presented in this paper are not publicly available at this time but may be obtained from the authors upon reasonable request. 

See Supplementary 1 and 2 for supporting contents at \url{https://doi.org/10.1364/BOE.509763}.
%%%%% References %%%%%

\section*{Published version and supplementary materials}
This manuscript is now published in Biomedical Optics Express as 
L.\@ Zhu, S.\@ Makita, J.\@ Tamaoki, Y.\@ Zhu, P.\@ Mukherjee, Y.\@ Lim, M.\@ Kobayashi, and Y.\@ Yasuno, ``Polarization-artifact reduction and accuracy improvement of Jones-matrix polarization-sensitive optical coherence tomography by multi-focus-averaging based multiple scattering reduction,'' \textit{Biomed. Opt. Express} {\bf 15}, 256--276 (2024), \url{https://doi.org/10.1364/BOE.509763}.

\bibliography{reference.bib} 

\end{document}